\documentstyle[12pt]{article}

\def\hybrid{\topmargin -20pt	\oddsidemargin 0pt
	\headheight 0pt	\headsep 0pt
	\textwidth 6.25in	
	\textheight 9.5in	
	\marginparwidth .875in
	\parskip 5pt plus 1pt	\jot = 1.5ex}

\hybrid

\def\baselinestretch{1.2}

\catcode`\@=11

\def\marginnote#1{}
\newcount\hour
\newcount\minute
\newtoks\amorpm
\hour=\time\divide\hour by60
\minute=\time{\multiply\hour by60 \global\advance\minute by-\hour}
\edef\standardtime{{\ifnum\hour<12 \global\amorpm={am}%
	\else\global\amorpm={pm}\advance\hour by-12 \fi
	\ifnum\hour=0 \hour=12 \fi
	\number\hour:\ifnum\minute<10 0\fi\number\minute\the\amorpm}}
\edef\militarytime{\number\hour:\ifnum\minute<10 0\fi\number\minute}

\def\draftlabel#1{{\@bsphack\if@filesw {\let\thepage\relax
   \xdef\@gtempa{\write\@auxout{\string
      \newlabel{#1}{{\@currentlabel}{\thepage}}}}}\@gtempa
   \if@nobreak \ifvmode\nobreak\fi\fi\fi\@esphack}
	\gdef\@eqnlabel{#1}}
\def\@eqnlabel{}
\def\@vacuum{}
\def\draftmarginnote#1{\marginpar{\raggedright\scriptsize\tt#1}}

\def\draft{\oddsidemargin -.5truein
	\def\@oddfoot{\sl preliminary draft \hfil
	\rm\thepage\hfil\sl\today\quad\militarytime}
	\let\@evenfoot\@oddfoot	\overfullrule 3pt
	\let\label=\draftlabel
	\let\marginnote=\draftmarginnote
   \def\@eqnnum{(\theequation)\rlap{\kern\marginparsep\tt\@eqnlabel}%
\global\let\@eqnlabel\@vacuum}  }

\def\preprint{\twocolumn\sloppy\flushbottom\parindent 2em
	\leftmargini 2em\leftmarginv .5em\leftmarginvi .5em
	\oddsidemargin -.5in	\evensidemargin -.5in
	\columnsep .4in	\footheight 0pt
	\textwidth 10.in	\topmargin  -.4in
	\headheight 12pt \topskip .4in
	\textheight 6.9in \footskip 0pt
	\def\@oddhead{\thepage\hfil\addtocounter{page}{1}\thepage}
	\let\@evenhead\@oddhead	\def\@oddfoot{}	\def\@evenfoot{} }

\def\numberbysection{\@addtoreset{equation}{section}
	\def\theequation{\thesection.\arabic{equation}}}

\def\underline#1{\relax\ifmmode\@@underline#1\else
	$\@@underline{\hbox{#1}}$\relax\fi}

\def\titlepage{\@restonecolfalse\if@twocolumn\@restonecoltrue\onecolumn
     \else \newpage \fi \thispagestyle{empty}\c@page\z@
	\def\thefootnote{\fnsymbol{footnote}} }

\def\endtitlepage{\if@restonecol\twocolumn \else \newpage \fi
	\def\thefootnote{\arabic{footnote}}
	\setcounter{footnote}{0}}  

\catcode`@=12
\relax

\def\figcap{\section*{Figure Captions\markboth
	{FIGURECAPTIONS}{FIGURECAPTIONS}}\list
	{Figure \arabic{enumi}:\hfill}{\settowidth\labelwidth{Figure
999:}
	\leftmargin\labelwidth
	\advance\leftmargin\labelsep\usecounter{enumi}}}
 \relax
\def\tablecap{\section*{Table Captions\markboth
	{TABLECAPTIONS}{TABLECAPTIONS}}\list
	{Table \arabic{enumi}:\hfill}{\settowidth\labelwidth{Table
999:}
	\leftmargin\labelwidth
	\advance\leftmargin\labelsep\usecounter{enumi}}}
 \relax
\def\reflist{\section*{References\markboth
	{REFLIST}{REFLIST}}\list
	{[\arabic{enumi}]\hfill}{\settowidth\labelwidth{[999]}
	\leftmargin\labelwidth
	\advance\leftmargin\labelsep\usecounter{enumi}}}
 \relax
\makeatletter
\newcounter{pubctr}
\def\publist{\@ifnextchar[{\@publist}{\@@publist}}
\def\@publist[#1]{\list
	{[\arabic{pubctr}]\hfill}{\settowidth\labelwidth{[999]}
	\leftmargin\labelwidth
	\advance\leftmargin\labelsep
	\@nmbrlisttrue\def\@listctr{pubctr}
	\setcounter{pubctr}{#1}\addtocounter{pubctr}{-1}}}
\def\@@publist{\list
	{[\arabic{pubctr}]\hfill}{\settowidth\labelwidth{[999]}
	\leftmargin\labelwidth
	\advance\leftmargin\labelsep
	\@nmbrlisttrue\def\@listctr{pubctr}}}
 \relax
\makeatother
\newskip\humongous \humongous=0pt plus 1000pt minus 1000pt

\newif\ifdtup

\relax

\def\s{\sigma}
\def\thefootnote{\fnsymbol{footnote}}
\def\be{\begin{equation}}
\def\ee{\end{equation}}
\def\ba{\begin{eqnarray}}
\def\ea{\end{eqnarray}}

\def\th{\theta}

\def\l{\lambda}

\def\e{\epsilon}
\begin{document}
\renewcommand{\theequation}{\thesection.\arabic{equation}}
\newcommand{\beq}{\begin{equation}}
\newcommand{\eeq}[1]{\label{#1}\end{equation}}
\newcommand{\ber}{\begin{eqnarray}}
\newcommand{\eer}[1]{\label{#1}\end{eqnarray}}
\begin{titlepage}
\begin{center}

\hfill CERN--TH/96--16\\
\hfill ENSLAPP--A--579/96\\
\hfill THU--96/05\\
\hfill hep--th/9604003\\

\vskip .2in

{\large \bf TODA FIELDS OF SO(3) HYPER--KAHLER METRICS \\
AND FREE FIELD REALIZATIONS}

\vskip 0.4in

{\bf Ioannis Bakas}
\footnote{Permanent address: Department of Physics, University of
Patras,
26110 Patras, Greece}
\footnote{e--mail address: BAKAS@SURYA11.CERN.CH,
BAKAS@LAPPHP8.IN2P3.FR}\\
\vskip .1in

{\em Theory Division, CERN, 1211 Geneva 23, Switzerland, and\\
Laboratoire de Physique Theorique ENSLAPP, 74941 Annecy-le-Vieux,
France}
\\

\vskip .3in

{\bf Konstadinos Sfetsos}
\footnote{e--mail address: SFETSOS@FYS.RUU.NL}\\
\vskip .1in

{\em Institute for Theoretical Physics, Utrecht University\\
     Princetonplein 5, TA 3508, The Netherlands}\\

\vskip .1in

\end{center}

\vskip .7in

\begin{center} {\bf ABSTRACT } \end{center}
\begin{quotation}\noindent
The Eguchi--Hanson, Taub--NUT and Atiyah--Hitchin metrics are the
only
complete non--singular $SO(3)$--invariant hyper--Kahler metrics in
four
dimensions. The presence of a rotational $SO(2)$ isometry allows for
their unified treatment based on solutions of the 3--dim continual
Toda equation. We determine the Toda potential in each case and
examine
the free field realization of the corresponding solutions, using
infinite
power series expansions. The Atiyah--Hitchin metric exhibits some
unusual features attributed to topological properties
of the group of area preserving diffeomorphisms. The construction of
a descending series of $SO(2)$-invariant 4--dim regular hyper--Kahler
metrics
remains an interesting question.
\end{quotation}
\vskip .2cm
March 1996\\
\end{titlepage}
\vfill
\eject

\def\baselinestretch{1.2}
\baselineskip 16 pt
\setcounter{section}{0}

\section{\bf Introduction}
\noindent
Hyper--Kahler manifolds in four dimensions have been studied
extensively
in general relativity over the last twenty years, in connection with
the
theory of gravitational instantons [1], and their algebraic
generalizations
through Penrose's non--linear graviton theory [2] and the heavenly
equations [3].
They have also been of
primary interest in supersymmetric field theories [4], where the
presence
of an extended $N=4$ supersymmetry imposes powerful constraints that
account
for various non--renormalization theorems, and their finiteness at
the
ultra--violet. More recently it was found that moduli spaces arising
in
topological field theories and other areas of mathematical physics
possess
hyper--Kahler structures, a notable example being the moduli space of
BPS
magnetic monopoles [5]. Similar structures are also encountered in
superstring
theory, and in many other moduli problems of physical
interest in quantum gravity.

The classification of all complete, regular, hyper--Kahler manifolds
remains an open question to this date, and even for some known
examples
the explicit form of the metric has been difficult or impossible to
determine so far. The dimensionality and the boundary conditions
imposed
on the manifold are clearly very important for such an investigation;
recall that in four dimensions the only compact (simply connected)
space of this type is
$K_{3}$, while there is an infinite class of non--compact
hyper--Kahler
manifolds. It is a well--known fact that all asymptotically locally
Euclidean (ALE) hyper--Kahler 4--manifolds admit an A--D--E
classification [6],
where the boundary at infinity is $S^{3} / R_{k}$ with $R_{k}$ being
any
discrete subgroup of $SU(2)$. Similar results exist for
asymptotically
locally flat (ALF) hyper--Kahler 4--manifolds, although a complete
catalogue is not yet available for them.

In this paper we consider the algebraic description of 4--dim
non--compact
hyper--Kahler manifolds that possess (at least) one abelian isometry
and
address the issue of their classification, without really focusing on
any
particular applications. The translational or the
rotational character of the corresponding Killing vector
fields is a decisive factor in our investigation. A Killing vector
field
$K_{\mu}$ satisfies ${\nabla}_{( \nu} K_{\mu )} = 0$ by definition,
while
the self--duality of the anti--symmetric part ${\nabla}_{[ \nu}
K_{\mu ]}$
provides the relevant distinction [7]: $K_{\mu}$ will be
called translational if it satisfies the condition
\be
{\nabla}_{\nu} K_{\mu} = \pm {1 \over 2} \sqrt{\det g} ~
{{\epsilon}_{\nu \mu}}^{\kappa \lambda} {\nabla}_{\kappa} K_{\lambda}
{}~,
\ee
otherwise it will be called rotational Killing vector field. The
latter
is also called non--tri--holomorphic to
distinguish from the former whose action is holomorphic with respect
to
all three Kahler structures [8]. The $\pm$ sign in (1.1)
is chosen according to the self--dual or the anti--self--dual nature
of
the underlying 4--dim metric $g_{\mu \nu}$.

We consider 4--dim metrics of the general form
\be
ds^{2} = V(d \tau + {\omega}_{i} d x^{i})^{2} + V^{-1} {\gamma}_{ij}
d x^{i} d x^{j} ~,
\ee
where $V$, ${\omega}_{i}$ and ${\gamma}_{ij}$ are all independent of
$\tau$,
and hence $K^{\mu} = (1, 0, 0, 0)$. It will be convenient in these
adapted
coordinates to use the notation $x = x^1$, $y = x^2$ and $z = x^3$.
For
translational Killing vector fields $K_{\mu}$, we may choose without
loss of
generality a coordinate system so that
\be
{\partial}_{i} (V^{-1}) = \pm {1 \over 2} {\epsilon}_{ijk}
({\partial}_{j} {\omega}_{k} - {\partial}_{k} {\omega}_{j}) ~, ~~~~
{\gamma}_{ij} = {\delta}_{ij} ~.
\ee
Then, for hyper--Kahler metrics (1.2), the self--duality (or
anti--self--duality)
condition is equivalent to the 3--dim Laplace equation [1]
\be
({\partial}_{x}^{2} + {\partial}_{y}^{2} + {\partial}_{z}^{2}) V^{-1}
= 0
\ee
with respect to the flat metric ${\delta}_{ij}$. Localized solutions
of the form
\be
V^{-1} = \epsilon + \sum_{i=1}^{n} {m \over \mid \vec{x} -
{\vec{x}}_{0, i} \mid}
\ee
describe non--singular multi--center 4--metrics with moduli
parameters $m$ and
${\vec{x}}_{0, i}$, provided that $\tau$ is periodic. The
(multi)--Eguchi--Hanson or
(multi)--Taub--NUT metrics are obtained for $\epsilon = 0$ or
$\epsilon \neq 0$
(in which case $\epsilon$ can be normalized to $1$) respectively;
$\epsilon = 0$
corresponds to the A--series of ALE spaces, while $\epsilon = 1$ to
ALF spaces [1].

For rotational Killing vector fields $K_{\mu}$, we may choose without
loss of
generality a coordinate system (1.2) so that
\be
V^{-1} = {\partial}_{z} \Psi ~, ~~~~ {\omega}_{1} = \mp
{\partial}_{y} \Psi ~,
{}~~~ {\omega}_{2} = \pm {\partial}_{x} \Psi ~, ~~~~ {\omega}_{3} = 0
\ee
and the diagonal $\gamma$--metric
\be
{\gamma}_{11} = {\gamma}_{22} = e^{\Psi} ~, ~~~~ {\gamma}_{33} = 1 ~,
\ee
in terms of a single scalar function $\Psi (x, y, z)$. Then, the
self--duality
(or anti--self--duality) condition of the metric is equivalent to the
3--dim
continual Toda equation [7]
\be
({\partial}_{x}^{2} + {\partial}_{y}^{2}) \Psi + {\partial}_{z}^{2}
(e^{\Psi}) = 0 ~.
\ee
This equation can be thought as arising in the large $N$ limit of the
2--dim Toda
theory based on the group $SU(N)$, where the Dynkin diagram becomes a
continuous
line parametrized by the third space variable $z$ [9, 10]. The main
problem that arises
in this context is to determine the solutions of (1.8) that
correspond to
complete, regular, 4--dim metrics, without worrying at the moment
about the
specific form of the boundary conditions. We note that although there
exists a
classification of the 4--dim hyper--Kahler metrics with (at least)
one
translational Killing symmetry, as was described above, a similar
result is not
yet available for metrics with rotational isometries.

There are a few results scattered in the literature about self--dual
solutions
with rotational Killing symmetries. Two of these examples are
provided by the
Eguchi--Hanson and the Taub--NUT metrics, which admit a much larger
group of
isometries, namely $SO(3) \times SO(2)$. For the first one, $SO(3)$
acts as a
translational isometry (in the sense of (1.1)) and $SO(2)$ as
rotational, while
the situation is reversed for the other [8]. In either case, there
exists a translational
Killing symmetry that accounts for their occurence in the known list
of ALE or ALF
spaces respectively. The only example known to this date that is
purely rotational,
without exhibiting any translational isometries, is the
Atiyah--Hitchin metric
on the moduli space $M_{2}^{0}$ of BPS $SU(2)$ monopoles of magnetic
charge $2$ [5]. These three examples
exhaust the list of $SO(3)$--invariant, complete, regular,
hyper--Kahler 4--metrics
[8], and they will provide the basis for our investigation in the
following.

It is worth stressing that not every solution of the continual Toda
equation (1.8)
has a good space--time interpretation, since the corresponding
metrics might be
incomplete with singularities. Furthermore, those ones that arise as
large $N$ limit solutions of the $SU(N)$ Toda theory do not
exhaust the list of all candidate metrics, in contrary to the naive
expectation
from large $N$ limit considerations. As we will see later, there are
solutions of the continual Toda equation that do not admit a free
field
specialization according to the standard group theoretical formula
[9].
We will point out the relevance of such solutions, showing
that the Atiyah--Hitchin metric is precisely of this type. Since the
Atiyah--Hitchin metric admits only rotational isometries, it may be
regarded as
the simplest representative from a series of purely rotational
hyper--Kahler 4--metrics. The explicit construction of a new series
of metrics will
certainly require a better and more systematic understanding of the
unusual features that arise in the continual limit of the Toda field
equations. The Atiyah--Hitchin metric, which has so far been regarded
as an isolated
example with respect to all known series, could provide the means for
exploring further this direction.

The present paper is an attempt to summarize the known results on the
subject,
including several new, and set up the framework for
exploring the construction of a Toda--like
series of hyper--Kahler metrics in four dimensions. In section 2, we
review
the classification of $SO(3)$--invariant hyper--Kahler metrics using
their
Bianchi IX formulation [11, 5, 8].
Although these metrics have more symmetry than it is
required on general grounds, they provide the only non--trivial
examples known to
this date with (at least) one rotational isometry. In section 3, we
formulate the
Eguchi--Hanson, Taub--NUT and Atiyah--Hitchin metrics in the adapted
rotational
coordinates (1.6), (1.7) and determine the Toda potential $\Psi$ in
each case.
In section 4, we consider the free field realization of solutions of
the Toda
field equations [12, 9]
and identify the free field configurations associated with the
Eguchi--Hanson and Taub--NUT metrics. In section 5 we
argue that a new class of solutions exist in the
continual limit, where the Atiyah--Hitchin metric fits quite
naturally.
Their occurence is attributed to topological properties of the group
of
area preserving diffeomorphisms,
whose commutation relations descibe $SU(\infty)$ as a continual
Lie algebra with distributional structure constants [9] (see also
[13]), but they
are not shared by $SU(N)$ for any finite $N$. Finally in section 6,
we present
our conclusions, and discuss further the framework for possible
generalizations.

\section{\bf SO(3) hyper-Kahler 4-metrics}
\setcounter{equation}{0}
\noindent
Metrics with $SO(3)$ isometry can be written using the Bianchi IX
formalism,
\be
ds^2 = f^{2}(t) dt^{2} + a^{2}(t) {\sigma}_{1}^{2} +
b^{2}(t) {\sigma}_{2}^{2} + c^{2}(t) {\sigma}_{3}^{2}  ~,
\ee
where
\ba
{\sigma}_{1} & = & {1 \over 2} (\sin \psi d \theta - \sin \theta \cos
\psi
d \phi ) ~ ,\nonumber\\
{\sigma}_{2} & = & - {1 \over 2} (\cos \psi d \theta + \sin \theta
\sin \psi
d \phi ) ~ , \\
{\sigma}_{3} & = & {1 \over 2} (d \psi + \cos \theta d \phi )
\nonumber
\ea
are the invariant 1--forms of $SO(3)$, which are normalized so that
${\sigma}_{i} \wedge {\sigma}_{j} = {1 \over 2} {\epsilon}_{ijk} d
{\sigma}_{k}$.
The Euler angles $\theta$, $\psi$, $\phi$ range as usual, while
$f(t)$ can
always be chosen as
\be
f(t) = {1 \over 2} a b c ~ ,
\ee
by appropriate reparametrization in $t$.

Self--dual metrics with $SO(3)$ isometry were studied some time ago
[11], where it
was found that the corresponding second--order differential equations
in $t$ can be
integrated once to yield the following first--order system:
\ba
2 ~ {a^{\prime} \over a} & = & b^2 + c^2 - 2{\lambda}_{1} bc - a^2 ~
,\nonumber\\
2 ~ {b^{\prime} \over b} & = & c^2 + a^2 - 2 {\lambda}_{2} ca - b^2 ~
,\\
2 ~ {c^{\prime} \over c} & = & a^2 + b^2 - 2 {\lambda}_{3} ab - c^2 ~
. \nonumber
\ea
The derivation of these equations assumes the choice (2.3) for the
coordinate $t$.
Also, depending on the conventions for having a self--dual or
anti--self--dual
metric, there is an ambiguity in the overall sign of (2.4); the two
cases are
related to each other by letting $t \rightarrow -t$.

We essentially have two distinct cases, depending on the values of
the
parameters ${\lambda}_{1}$, ${\lambda}_{2}$, ${\lambda}_{3}$. The
first case is
described by ${\lambda}_{1} = {\lambda}_{2} = {\lambda}_{3} = 0$ and
corresponds to the Eguchi--Hanson metric, while for
${\lambda}_{1} = {\lambda}_{2} = {\lambda}_{3} = 1$ we obtain the
Taub--NUT
metric. These two solutions, apart from an $SO(3)$ isometry, also
exhibit an
additional $SO(2)$ symmetry that arises from equating two of the
metric
coefficients, say $a = b$ [11].
A third possibility was found a few years later,
also having ${\lambda}_{1} = {\lambda}_{2} = {\lambda}_{3} = 1$,
but with unequal metric coefficients $a$, $b$, $c$ for generic values
of the
parameter $t$ [5]. This solution is known as the Atiyah--Hitchin
metric, and it has
been used for studying the geometry and
dynamics of $SU(2)$ monopoles with magnetic charge 2.
It was subsequently shown that these three cases provide the only
non--trivial
hyper--Kahler 4--metrics with $SO(3)$ isometry that are complete and
non--singular [8].

Before discussing explicit solutions,
we present the expressions for the three
Kahler forms for the general $SO(3)$--invariant
hyper--Kahler metric (2.1)--(2.4).
They are given by [14]
\ba
F_i = \left \{ \begin{array} {ccc}
 K_i~~ & {\rm if}~~ &  \l_1=\l_2=\l_3=0  \\
  C_{ij} K_j ~~&{\rm if}~~ & \l_1=\l_2=\l_3=1  \\
\end{array}
\right\} ~ ,
\ea
where
\be
K_i = 2 e_0 \wedge e_i +  \e_{ijk} e_j \wedge e_k ~ ,
\ee
with the tetrads defined as $e_0= f dt$, $e_1 = a \s_1$,
$e_2 = b \s_2$ and $e_3 = c \s_3$.
The matrix
$(C_{ij})$ defines the adjoint representation of $SO(3)$
\be
C_{ij}  = {1\over 2} Tr(\s_i g \s_j g^{-1})~ , ~~~~
g=e^{{i\over 2} \phi \s_3}
e^{{i\over 2} \th \s_2} e^{{i\over 2} \psi \s_3} ~ .
\ee
Of course, one should make a distinction between the invariant
1--forms of
$SO(3)$ and the Pauli matrices used above.

(i) \underline{Eguchi--Hanson metric} : The solution of the
differential equations
(2.4) with $a=b$ and ${\lambda}_{1} = {\lambda}_{2} = {\lambda}_{3} =
0$ can be
easily found. Letting $t \rightarrow -t$ for convenience, which
amounts to flipping
the sign in (2.4), we have
\be
a^2 = b^2 = m^2 \coth (m^2 t) ~, ~~~~ c^2 = {2 m^2 \over \sinh (2 m^2
t)} ~ ,
\ee
where $m$ is the moduli parameter of the Eguchi--Hanson metric. In
this case, the
metric coefficient $f(t)$ in (2.1) is given by (2.3). The more
standard description
of the Eguchi--Hanson metric follows by introducing $r$,
\be
r^2 = m^2 \coth (m^2 t) ~ ,
\ee
which yields the line element
\be
ds^2 = {dr^2 \over 1 - { \left( {m \over r} \right)}^{4}} +
r^2 \left( {\sigma}_{1}^{2} + {\sigma}_{2}^{2} +
\left( 1  - { \left( {m \over r} \right)}^{4} \right)
{\sigma}_{3}^{2} \right) ~ .
\ee
It is well--known that the $SO(3)$ isometry of this metric acts in a
translational
(or tri--holomorphic) way, while the Killing vector field $\partial /
\partial \psi$,
which arises from the equation $a = b$, acts as a rotational isometry
[8].

(ii) \underline{Taub--NUT metric} : The solution of the differential
equations (2.4)
with ${\lambda}_{1} = {\lambda}_{2} = {\lambda}_{3} = 1$ and $a = b$
is given by
\be
a^2 = b^2 = {1 + 4 m^2 t \over 4 m^2 t^2} ~ , ~~~~
c^2 = {4 m^2 \over 1 + 4 m^2 t} ~ ,
\ee
describing the Taub--NUT metric with a moduli parameter $m$. A more
conventional
description is obtained by introducing $r$,
\be
r = m + {1 \over 2 m t} ~ ,
\ee
which yields the line element
\be
ds^2 = {1 \over 4} {r + m \over r - m} dr^2 + (r^2 -
m^2)({\sigma}_{1}^{2} +
{\sigma}_{2}^{2}) + 4 m^2 {r - m \over r + m} {\sigma}_{3}^{2} ~ .
\ee
An alternative formulation is also given in terms of $\tilde{r} =
r-m$.
This metric also admits an additional isometry generated  by the
Killing vector field
$\partial / \partial \psi$, which now turns out to be translational.
On the
contrary, the $SO(3)$ isometry of the Taub--NUT metric is rotational
with respect to
all of its generators.

(iii) \underline{Atiyah--Hitchin metric} : The Taub--NUT metric is
the only
globally defined solution of the system (2.4) with
${\lambda}_{1} = {\lambda}_{2} = {\lambda}_{3} = 1$, provided that
$a$, $b$, $c$ are
all positive. Global considerations in this case force the existence
of an additional
$SO(2)$ isometry. The Atiyah--Hitchin metric on the other hand has $c
< 0$,
and moreover $a$, $b$, $c$ are all
different from each other for generic values of $t$. Equivalently one
may
consider all $a$, $b$ and $c$ positive and choose
${\lambda}_{1} = {\lambda}_{2} = -1$ and ${\lambda}_{3} = 1$.

The metric components are given in terms of the complete elliptic
integrals of
the first and second kind
\be
K(k) = \int_{0}^{\pi / 2} {d \gamma \over \sqrt{1 - k^2 {\sin}^2
\gamma}} ~ ,
{}~~~ E(k) = \int_{0}^{\pi / 2} \sqrt{1 - k^2 {\sin}^{2} \gamma} ~ d
\gamma ~ .
\ee
In particular, using the parametrization
\be
ds^2 = {1 \over 4} a^2 b^2 c^2 \left({dk \over k {k^{\prime}}^2 K^2}
\right)^2
+ a^2 (k) {\sigma}_{1}^2 + b^2 (k) {\sigma}_{2}^2 + c^2 (k)
{\sigma}_{3}^{2} ~ ,
\ee
the solution obtained by Atiyah and Hitchin is given as function of
$k$ [5, 15],
\ba
ab & = & - K(k) (E(k) - K(k)) ~ , \nonumber\\
bc & = & - K(k) (E(k) - {k^{\prime}}^{2} K(k)) ~ , \\
ac & = & - K(k) E(k) ~ . \nonumber
\ea
Here, $0 < k < 1$ and ${k^{\prime}}^2 = 1 - k^2$.

In the limit $k \rightarrow 1$, the metric becomes exponentially
close to
Taub--NUT. This can be easily seen if we define
\be
k^{\prime} = \sqrt{1 - k^2} \simeq 4 \exp \left( {1 \over \gamma}
\right)
\ee
for $\gamma \rightarrow 0^{-}$. It follows from standard expansions
of the
elliptic integrals [16] that in this vicinity
\be
a^2 \simeq b^2 \simeq {1 + \gamma \over {\gamma}^2} ~ , ~~~~
c^2 \simeq {1 \over 1 + \gamma} ~ ,
\ee
which is the Taub--NUT configuration (2.11) with
$t = \gamma$ and $m^2 = 1 / 4$; in fact, as
$k \rightarrow 1$, one obtains the Taub--NUT metric with a negative
mass
parameter $m = - 1 / 2$.

The coordinate $t$ in the parametrization (2.1) and (2.3) is given by
the
change of variables
\be
t = - {2 K(k^{\prime}) \over \pi K(k)} ~ ,
\ee
up to an additive constant.
Yet another convenient parametrization of the metric components is
given
in terms of the coordinate
\be
r = 2 K(k) ~ , ~~~~~ \pi < r < \infty ~ ,
\ee
in which case the Taub--NUT limit is attained asymptotically as
$r \rightarrow \infty$. In this parametrization it becomes more clear
that
asymptotically the Taub-NUT parameter is negative [5].

The generators of the $SO(3)$ isometry act as rotational vector
fields in this
case [8]
and, therefore, this metric provides the only known example of a
purely
rotational hyper--Kahler 4--metric. The Atiyah--Hitchin metric has
many
similarities with the Taub--NUT, but the absence of an additional
(translational)
isometry in the former has profound implications in monopole physics,
where slowly
moving monopoles can be converted into dyons [5].

We also note for completeness that there have
been attempts in the past to construct exact self--dual metrics of
Bianchi
IX type with all directions unequal, which generalize the
Eguchi--Hanson
metric. These solutions have ${\lambda}_{1} = {\lambda}_{2} =
{\lambda}_{3} = 0$,
and although asymptotically
Euclidean they contain singularities [11, 17]. A particularly simple
parametrization
of them is given by
\be
ds^2 = {1 \over 4} P^{- 1/2} dx^2 + P^{1/2} \left(
{{\sigma}_{1}^{2} \over x - x_1} +
{{\sigma}_{2}^{2} \over x - x_2} +
{{\sigma}_{3}^{2} \over x - x_3} \right) ~ ,
\ee
where
\be
P(x) = (x - x_{1}) (x - x_{2}) (x - x_{3})
\ee
and $x_1$, $x_2$, $x_3$ are the relevant moduli parameters. We also
mention that
the $SO(3)$ isometry acts as a translational symmetry in this case;
the
coordinate transformation that brings (2.21) into the form
(1.2)--(1.4) is also known
[17]. In this regard, the Atiyah--Hitchin
metric is the unique non--singular solution with $SO(3)$ symmetry
having
unequal coefficients $a$, $b$ and $c$.

\section{\bf Toda potential of the metrics}
\setcounter{equation}{0}
\noindent
There is an alternative algebraic description of the translational or
the
rotational character of a Killing vector field based on the notion of
the
nut potential. Using the adapted coordinate system (1.2) for a 4--dim
metric with a Killing vector field $\partial / \partial \tau$, we
define
the nut potential $b_{nut}$ associated with this isometry as follows
[1]:
\be
{\partial}_{i} b_{nut} = {1 \over 2} V^2
\sqrt{\det \gamma} ~ {{\epsilon}_{i}}^{jk}
({\partial}_{j} {\omega}_{k} - {\partial}_{k} {\omega}_{j}) ~ .
\ee
The compatibility condition for the system (3.1) is provided by the
vacuum
Einstein equations for the metric (1.2), and hence $b_{nut}$
exists only on--shell.

Consider now the case of self--dual or anti--self--dual metrics and
define
respectively the field
\be
S_{\pm} = b_{nut} \pm V ~ .
\ee
We also define the characteristic quadratic quantity
\be
\Delta S_{\pm} = {\gamma}^{ij} ({\partial}_{i} S_{\pm})
({\partial}_{j} S_{\pm}) ~ ,
\ee
which is clearly $\geq 0$. It is a well--known theorem [7] that for
translational
Killing vector fields (1.1) $\Delta S_{\pm} = 0$, while for
rotational
$\Delta S_{\pm} > 0$. In the former case this means that $S_{\pm} =
0$, up to
an additive constant, while in the latter $S_{\pm}$ is a coordinate
dependent
configuration. In fact, for rotational isometries, the passage to the
special
coordinate system (1.6)--(1.8) can be achieved by considering
$S_{\pm}$ as
the $z$--coordinate, since in this case $\Delta S_{\pm} = 1$. In
general
we may choose
\be
z = {S_{\pm} \over \sqrt{\Delta S_{\pm}}} ~ ,
\ee
and then determine the transformation of the remaining coordinates
that
brings the metric into the desired Toda form. This is precisely the
prescription that will be followed to determine the Toda potential
$\Psi$ of the three $SO(3)$--invariant hyper--Kahler metrics in
question.
The computation of the nut potential is clearly very important for
performing the required coordinate transformation in each case.

We choose our conventions in the following so that the relevant
variable
is the field $S_{+}$ instead of $S_{-}$.

The Toda frame formulation of the $SO(3)$ hyper--Kahler metrics is
also
equivalent to the problem of determining the explicit form of the
Kahler
potential and the corresponding Kahler coordinates, using the method
of
Boyer and Finley [7]. Moreover, the three complex structures form an
$SO(2)$--doublet and a singlet with respect to every rotational
isometry.
The three Kahler forms can be written explicitly in the Toda frame
formulation
[18]; we have
\be
\left(\begin{array}{c}
F_{1} \\
      \\
F_{2} \end{array} \right) = e^{{1 \over 2} \Psi}
\left(\begin{array}{cr}
\cos {\tau \over 2} & \sin {\tau \over 2} \\
                    &                     \\
\sin {\tau \over 2} & - \cos {\tau \over 2} \end{array} \right)
\left(\begin{array}{c}
f_{1} \\
      \\
f_{2} \end{array} \right)
\ee
for the doublet, where
\ba
f_{1} & = & (d \tau + {\omega}_{2} dy) \wedge dx - V^{-1}
dz \wedge dy ~ , \nonumber \\
f_{2} & = & (d \tau + {\omega}_{1} dx) \wedge dy + V^{-1} dz \wedge
dx ~ ,
\ea
while for the singlet we have
\be
F_{3} = (d \tau + {\omega}_{1} dx + {\omega}_{2} dy) \wedge dz +
V^{-1} e^{\Psi} dx \wedge dy ~ .
\ee
We also note that all generators of the $SO(3)$ isometry act either
in a
translational or in a rotational fashion, in which case the three
complex
structures form either singlets or an $SO(3)$--triplet [8].

(i) \underline{Eguchi--Hanson metric} : Starting from the
Eguchi--Hanson metric
in the form (2.10), we note that $\partial / \partial \psi$ is a
Killing vector
field thanks to the equality $a^2 = b^2 = r^2$. It is relatively easy
to
calculate the nut potential associated with this isometry, using the
defining
relations (3.1). In this parametrization we find
\be
V = {1 \over 4} r^2 \left( 1 - {m^4 \over r^4} \right) ~ , ~~~~
b_{nut} = {1 \over 4} r^2 \left( 1 + {m^4 \over r^4} \right) ~ ,
\ee
and hence $S_{+} = b_{nut} + V = r^2 / 2$. The result is consistent
with the fact
that $\partial / \partial \psi$ is rotational. We also find that
$\Delta S_{+} = 4$ and therefore, according to the previous
discussion, the
first step in transforming the Eguchi--Hanson metric into the Toda
frame
consists of the choice $z = r^2 / 4$. The remaining change of
coordinates,
as well as the identification of the Toda potential, is easily done
by rescaling the metric (2.10) with an overall factor of 2. The
complete
transformation can be summarized as follows:
\ba
x & = & 2 \sqrt{2} \cos \phi \tan {\theta \over 2} ~ , ~~~~
y = 2 \sqrt{2} \sin \phi \tan {\theta \over 2} ~ , \nonumber \\
z & = & {1 \over 4} r^2 ~ , ~~~~~ \tau = 2 (\psi + \phi) ~ .
\ea
The $\phi$--shift in the definition of $\tau$ is used in order to set
${\omega}_{3} = 0$ for convenience. Then, the metric assumes the form
(1.2), with (1.6) and (1.7) satisfied, where
\be
e^{\Psi} = {z^2 - {\alpha}^2 \over 2 \left(1+ {1 \over 8} (x^2 + y^2)
\right)^2} ~ , ~~~~ z^2 \geq {\alpha}^2 ~ ,
\ee
with $4 \alpha = m^2$. The potential $\Psi$ we have determined
clearly obeys
the continual Toda equation (1.8) (see also [7, 10]).

The Killing vector field $\partial / \partial \phi$ is translational,
as it
can be verified directly by computing the corresponding nut
potential, which
yields $S_{+} = 0$. The explicit transformation to the translational
frame
(1.3)--(1.5) is known in this case [19]; it simply reads,
\be
\vec{x} = {1 \over 8} \left( \sqrt{r^4 - m^4 }\sin \theta \cos \psi ,
{}~
\sqrt{r^4 - m^4 } \sin \theta \sin \psi , ~ r^2 \cos \theta \right) ,
{}~~~~
\tau = 2 \phi ,
\ee
with
\ba
V^{-1} & = & {1 \over r_{+}} + {1 \over r_{-}} ~ ; ~~~~
r_{\pm}^{2} = x^2 + y^2 + z_{\pm}^2 ~ , ~~~~ z_{\pm} =
z \pm {m^2 \over 8} ~ , \nonumber\\
\vec{\omega} & = & {1 \over x^2 + y^2} \left( {z_{+} \over r_{+}} +
{z_{-} \over r_{-}} \right) (-y , ~ x , ~ 0) ~ .
\ea

(ii) \underline{Taub--NUT metric} : Here, the Killing vector field
$\partial / \partial \psi$ is translational, and the explicit
transformation
to the coordinate system (1.3)--(1.5) is given by
\be
\vec{x} = (r - m) (\sin \theta \cos \phi , ~ \sin \theta \sin \phi ,
{}~
\cos \theta ) ~ , ~~~~
\tau = {m \over 2} \psi ~ ,
\ee
with
\be
V^{-1} = {1 \over 4} \left( 1 + {2m \over \mid \vec{x} \mid } \right)
,
{}~~~ \vec{\omega} = {m z \over 2 \mid \vec{x} \mid (x^2 + y^2 )}
(-y , ~ x , ~ 0) ~.
\ee

The isometry generated by
$\partial / \partial \phi$, however, is rotational and it will be
used to determine the Toda potential of the Taub--NUT metric.
Starting
from the Bianchi IX formulation (2.1)--(2.4), we calculate
first the nut potential of the corresponding isometry. The explicit
form
(2.11) of the coefficients $a = b$ and $c$ is not required for the
computation,
as they may be substituted only in the final expression. We obtain,
\be
b_{nut} = {1 \over 4} c \left(2a - c - (a - c) {\sin}^2 \theta
\right) ,
{}~~~
V = {1 \over 4} \left(c^2 + (a^2 - c^2) {\sin}^2 \theta \right) ~ .
\ee
We also consider $S_{+} = b_{nut} + V$ and find $\Delta S_{+} = 4$.
Hence, we
define the $z$--coordinate as
\be
z = {1 \over 8} a \left(2c + (a - c) {\sin}^2 \theta \right) ~ \equiv
{}~
{1 \over 4t} \left(1 + {1 \over 8 m^2 t} {\sin}^2 \theta \right) ~ .
\ee
Since the metric ${\gamma}_{ij}$ is diagonal, the $x$ and $y$
coordinates
can be easily found to be
\be
x = \psi ~ , ~~~~ y = {c - a \over c} \cos \theta +
\log \left( \tan {\theta \over 2} \right) ~ \equiv ~
-{1 \over 4 m^2 t} {\cos} \theta + \log \left(\tan {\theta \over 2}
\right) ~ ,
\ee
and moreover,
\be
\tau = 2 \phi ~ .
\ee
Like the Eguchi--Hanson metric, it is also convenient
here to rescale the Bianchi IX
metric with a factor of 2, and in
the process we find the
Toda potential (see also [20]),
\be
e^{\Psi} = {1 \over 16} a^2 c^2 {\sin}^2 \theta ~ \equiv ~
{1 \over 16 t^2} {\sin}^2 \theta ~ .
\ee

We note that the Toda potential of the Taub--NUT metric is
independent of $x$,
satisfying a reduced 2--dim continual Toda equation. This is merely a
reflection
of the additional $SO(2)$ isometry that arises from the
identification
$a^2 = b^2$ in this case. As we will see later, the Atiyah--Hitchin
metric,
which generalizes Taub--NUT with $a^2 \neq b^2$, does not share this
feature.
Also, one of the main differences with the Eguchi--Hanson result
is that the $\Psi$ above can
only be implicitly written as a function of $y$ and $z$, because the
coordinate transformation that is involved,
$(t, \theta) \rightarrow (y, z)$, can not be inverted in closed form.
For this reason we expect that the Toda potential of the
Atiyah--Hitchin
metric will be an even more complicated implicit function of the
coordinates
$x$, $y$ and $z$.

It is also worth clarifying the meaning of the Toda frame in the case
that
the moduli parameter of the Eguchi--Hanson or the Taub--NUT metric is
chosen so that the flat space limit is attained. Taking
$m \rightarrow 0$ for the Eguchi--Hanson metric (2.10) or $m
\rightarrow \infty$
for the Taub--NUT metric (2.11) we obtain the flat space metric
\be
ds^2 = dR^2 + R^2 ({\sigma}_{1}^2 + {\sigma}_{2}^2 + {\sigma}_{3}^2 )
{}~ ,
\ee
with $R = r$ and $R^2 = 1/t$ respectively. In these limiting cases
both
vector fields $\partial / \partial \psi$ and $\partial / \partial
\phi$
generate translational isometries: although $S_{+}$ remains
non--trivial,
$S_{-} \rightarrow 0$ in both cases, and since flat space is
self--dual as well as
anti--self--dual the rotational character of $\partial / \partial
\psi$
or $\partial / \partial \phi$ disappears. Then, the transformation to
the
Toda frame can only be regarded as a change of
coordinates in flat space,
with no reference to any rotational isometries. Note that in this
limit the
change of variables (3.16) and (3.17) for Taub--NUT can be inverted
so that
\be
z = {1 \over 4 t} ~ , ~~~~ {\sin} \theta = {1 \over \cosh y} ~ , ~~~~
e^{\Psi} = {z^2 \over {\cosh}^2 y} ~ .
\ee
This example, together with the $\alpha \rightarrow 0$ limit of
(3.10)
demonstrate explicitly that not every solution of the continual Toda
equation
corresponds to rotational isometries. Furthermore, algebraically
inequivalent solutions could yield 4--dim hyper--Kahler metrics that
are
simply related by diffeomorphisms.

(iii) \underline{Atiyah--Hitchin metric} : The starting point here is
also the
Bianchi IX formulation (2.1)--(2.4) of the metric. The isometry
generated by the
Killing vector field $\partial / \partial \phi$ is again rotational
and the
corresponding nut potential can be calculated without knowing the
explicit
form of the coefficients $a$, $b$ and $c$, but only their
differential equations
in $t$. These coefficients are implicit
functions of $t$ through the defining relation (2.19). We have
determined
that
\ba
b_{nut} & = & {1 \over 4} \left(c(a+b-c) - (a-c) (a+c-b) {\sin}^2
\theta +
(a-b) (a+b-c) {\sin}^2 \theta {\sin}^2 \psi \right) ~ , \nonumber \\
V & = & {1 \over 4} \left(c^2 + (a^2 - c^2) {\sin}^2 \theta
- (a^2 - b^2) {\sin}^2 \theta {\sin}^2 \psi \right) ~ .
\ea
The field $S_{+} = b_{nut} + V$ satisfies $\Delta S_{+} = 4$ as
before, and hence
one of the coordinates in the desired Toda frame is
\ba
z & = & {1 \over 8} \left(c(a+b) + b(a-c) {\sin}^2 \theta -
c(a-b) {\sin}^2 \theta {\sin}^2 \psi \right) \nonumber\\
& \equiv & {1 \over 8} K^2 (k) \left( k^2 {\sin}^2 \theta +
{k^{\prime}}^2 (1 + {\sin}^2 \theta {\sin}^2 \psi ) -
2 {E(k) \over K(k)} \right) .
\ea
Notice that if $a=b$ the above expressions reduce to (3.15) and
(3.16)
respectively for the Taub--NUT metric.

The appropriate transformation for determining $x$, $y$ and $\tau$
(with the
choice ${\omega}_{3} = 0$) is rather difficult to perform, but
fortunately
Olivier has already provided the result [15]. It is convenient to
rescale
the metric by a factor of 2 and introduce the quantities
\be
\mu = {2ab - c(a+b) \over c(b-a)} \equiv {1 + k^2 \over 1 - k^2} ~ ,
{}~~~~
\nu = 2 \left(\log \left( \tan {\theta \over 2} \right) + i \psi
\right) ~ .
\ee
Then, using our conventions, and substituting the expressions (2.16)
for the
metric coefficients, one finds that the coordinates $y$ and $x$
coincide with the
real and the imaginary parts of the complex variable
\be
y + ix = K(k) \sqrt{1 + {k^{\prime}}^2 {\sinh}^2 {\nu \over 2}}
\left( \cos \theta + {\tanh {\nu \over 2} \over K(k)}
\int_{0}^{\pi / 2} d \gamma
{\sqrt{1 - k^2 {\sin}^2 \gamma} \over 1 - k^2 {\tanh}^2 {\nu \over 2}
{\sin}^2
\gamma} \right) ~ ,
\ee
and also
\be
\tau = 2 \left(\phi + \arg \left(1 + {k^{\prime}}^2 {\sinh}^2 {\nu
\over 2}
\right) \right) ~ .
\ee
Furthermore, the Toda potential in this case turns out to be
\be
e^{\Psi} = {1 \over 32} c (b-a) {\sin}^2 \theta \mid \mu + \cosh \nu
\mid \equiv {1 \over 16} K^2 (k) {\sin}^2 \theta \mid 1 + 
{k^{\prime}}^2 {\sinh}^2 {\nu \over 2} \mid ~ .
\ee

Notice that as $a \rightarrow b$ the Toda potential of the metric
becomes
\be
e^{\Psi} \rightarrow {1 \over 16} a (a-c) {\sin}^2 \theta ~ ,
\ee
which at first sight seems to disagree with the expression (3.19) for
the
Taub--NUT metric. There is no contradiction, however, because it can
be
readily verified using the Taub--NUT coefficients (2.11) that
$a^2 c^2 = a (a-c)$ provided that the moduli parameter $m^2 = 1/4$.
This is precisely the case of interest in the limit,
where the Atiyah--Hitchin metric can be approximated with
the Taub--NUT
metric. Similarly, expanding $K(k)$ as $k \rightarrow 1$,
and using the parametrization (2.17), we find that
Olivier's coordinate transformation yields the result we have already
described for the Taub--NUT metric. In the physical context of
monopole dynamics, the limit $k \rightarrow 1$ corresponds to
configurations where the two monopoles are very far away from each
other.

Another interesting limit is $k \rightarrow 0$, for which we find
\be
e^{\Psi} \rightarrow {{\pi}^2 \over 64} (1 - {\sin}^2 \theta
{\sin}^2 \psi ) \equiv - {1 \over 2} z ~ ,
\ee
and
\be
y \rightarrow 0 ~ , ~~~~ x \rightarrow {\pi \over 2} \sin \theta \sin
\psi ~.
\ee
In this limit, the 3--dim orbit of $SO(3)$ collapses to a 2--sphere,
and
there is a coordinate singularity of bolt type [5]. In the physical
context of monopole dynamics, this limit corresponds to
configurations
where the two monopoles coincide.

\section{\bf Free field realizations}
\setcounter{equation}{0}
\noindent
We are primarily interested in understanding the algebraic
differences
of the Atiyah--Hitchin metric from the other two metrics,
in view of possible
generalizations that this configuration may have in purely rotational
hyper--Kahler geometry. The results we have included so far are not
very illuminating in their present form, although they have been
described directly in the Toda frame that unifies all metrics in
question. Next, we will consider the free field realization of the
general solution of the continual Toda equation, as it is derived by
taking the large $N$ limit of the 2--dim $SU(N)$ Toda systems [9].
The Toda potential of the three metrics will be considered from this
particular point of view, which turns out to be most
appropriate for understanding the qualitative differences we are
looking for the Atiyah--Hitchin metric.

Consider the continual Toda equation (1.8) and supply the
non--linearity
with a parameter ${\lambda}^{2}$,
\be
({\partial}_{x}^2 + {\partial}_{y}^2 ) \Psi +
{\lambda}^2 {\partial}_{z}^2 (e^{\Psi}) = 0 ~ .
\ee
A solution admits a free field realization if $\Psi$ has a power
series
expansion,
\be
\Psi = {\Psi}_{0} + {\lambda}^2 {\Psi}_{1} + {\lambda}^4 {\Psi}_{2}
+ \cdots ~ ,
\ee
where ${\Psi}_{0}$ is $z$--dependent and
satisfies the 2--dim free field equation
$({\partial}_{x}^2 + {\partial}_{y}^2 ) {\Psi}_{0} = 0$. Then,
${\Psi}_{1}$, ${\Psi}_{2}$, $\dots$ can be determined recursively
order
by order in ${\lambda}^2$ from the field equation (4.1). In this
regard $\lambda \neq 0$ is a book--keeping parameter, since it can be
absorbed
by rescaling $x$ and $y$, or equivalently shifting $\Psi$ by
$2 \log \lambda$; it organizes the free field expansion of a
given solution and it should be set equal to $1$ at the end.
The advantage of this realization is that ${\Psi}_{0}$ has a much
simpler form compared to exact solutions of the full non-linear
problem.
For this reason we will search for the free field configurations
that correspond to the Eguchi--Hanson, Taub--NUT and Atiyah--Hitchin
metrics, in order to examine more transparently the differences of
the latter
from the former two solutions. Although the free field realization
of a given solution may not be uniquely determined, as it can be seen
in Liouville theory whose general solution is invariant under
fractional
$SL(2)$ transformations of the building holomorphic and
anti--holomorphic
field variables, the existence of free fields simplifies considerably
the
description of the solutions of the underlying non--linear problem.

For $SU(N)$ Toda theories all the solutions admit a free field
realization,
which can be systematically described using highest weight
representations.
It is convenient to introduce the system of Weyl generators
$H_{i}$, $X_{i}^{\pm}$ $(1 \leq i \leq N-1)$,
\be
[H_{i} , X_{j}^{\pm}] = \pm K_{ij} X_{j}^{\pm} ~ , ~~~~
[X_{i}^{+} , X_{j}^{-} ] = {\delta}_{ij} H_{j} ~ , ~~~~ [H_{i} ,
H_{j}] = 0 ~ ,
\ee
where $K$ is the Cartan matrix of $SU(N)$, and define matrices
$M_{\pm} (q_{\pm})$ as follows,
\be
{\partial}_{\pm} M_{\pm} (q_{\pm}) = M_{\pm} (q_{\pm})
\left( \lambda \sum_{i=1}^{N-1} e^{{\psi}_{i}^{\pm} (q_{\pm})}
X_{i}^{\pm}
\right) ; ~~~~~ q_{\pm} = {1 \over 2} (y \pm i x) ~ .
\ee
The fields ${\psi}_{i}^{\pm} (q_{\pm})$ are $N-1$ arbitrary
holomorphic and
anti--holomorphic functions so that
${\psi}_{i}^{+} (q_{+}) + {\psi}_{i}^{-} (q_{-})$ satisfy the 2--dim
free wave
equation. We also consider highest weight states $\mid j >$,
\be
X_{i}^{+} \mid j > = 0 ~ , ~~~~ < j \mid X_{i}^{-} = 0 ~ , ~~~~
H_{i} \mid j > = {\delta}_{ij} \mid j > ~ ,
\ee
and formulate the general solution of the 2--dim Toda field equation
\be
{\partial}_{+} {\partial}_{-} {\psi}_{i} (q_{+} , q_{-}) =
{\lambda}^{2} \sum_{j=1}^{N-1} K_{ij} e^{{\psi}_{j} (q_{+} , q_{-})}
{}~ .
\ee
It is given by the well--known expression [12],
\be
{\psi}_{i} (q_{+} , q_{-}) = {\psi}_{i}^{+} (q_{+}) + {\psi}_{i}^{-}
(q_{-})
- \sum_{j=1}^{N-1} K_{ij} \log < j \mid M_{+}^{-1} M_{-} \mid j > ~ ,
\ee
and hence it clearly admits a power series expansion in ${\lambda}^2$
around
the free field configurations ${\psi}_{i}^{+} (q_{+}) +
{\psi}_{i}^{-} (q_{-})$
that characterize every exact solution.

The matrices $M_{\pm}$ that follow by integration of (4.4) are
path--ordered
exponentials of $SU(N)$ Lie algebra elements and therefore they can
be
regarded as one--parameter family subgroups of the corresponding
gauge group
with values in $SU(N)$;
the free fields in their exponentiated form play the role of the
canonical
coordinates around the identity element of the group. Furthermore,
for $N$
finite, $\exp (- {\psi}_{i})$ admits a power series expansion that
terminates
at a finite order in $\lambda$, as it can be easily seen by
considering the
terms
\be
D_{j}^{\{i_{1} , \cdots , i_{n} ; i_{1}^{\prime} , \cdots ,
i_{n}^{\prime}\}}
= < j \mid X_{i_{1}}^{+} X_{i_{2}}^{+} \cdots X_{i_{n}}^{+}
X_{i_{n}^{\prime}}^{-} \cdots X_{i_{2}^{\prime}}^{-}
X_{i_{1}^{\prime}}^{-} \mid j >
\ee
in the power series expansion of the path--ordered exponentials in
(4.7); only
terms with $n < N$ are non--zero and contribute. The algebra of
Lie--Backlund
transformations is finite dimensional in this case, and it is not
possible to have
additional terms contributing to the general form of the
solutions. We also note that the solutions of the
Toda field equations have the following interpretation
in 2--dim Minkowski space, which is obtained by analytic continuation
$ix \rightarrow x$: the free field configurations
${\psi}_{i}^{\pm} (q_{\pm})$ are the boundary values of the Toda
field
${\psi}_{i} (q_{+} , q_{-})$ on the light--cone, and if the reference
light--cone point is
taken at $0$ the corresponding free fields have to be normalized by
substructing the constant value ${\psi}_{i} (0) / 2$. In 2--dim
Euclidean space,
which is relevant for our geometrical problem, the solution (4.7) of
the
boundary value problem for the Toda field equation is determined via
the
solution of the Laplace equation that corresponds to the free theory,
namely
the asymptotic values of the non--interacting free fields
${\psi}_{i}^{\pm} (q_{\pm})$.

For large $N$ we obtain the continual Toda equation as a limit,
where $\Psi (q_{+} , q_{-} , z)$ is a master field for all
${\psi}_{i} (q_{+} , q_{-})$, and
$K_{ij} \rightarrow - {\partial}_{z}^2 \delta (z - z^{\prime})$,
$\delta_{ij} \rightarrow \delta (z - z^{\prime})$. The introduction
of the
new coordinate $z$ turns summations over $i$ and $j$ into
integrations that can
be easily performed using the $\delta$--functions. Taking the
continual
version of the algebraic data (4.3)--(4.5), as it was done in [9,
10], we
obtain solutions of the continual Toda equation of the form
\be
\Psi(q_{+} , q_{-} , z) = {\Psi}_{0} (q_{+} , q_{-} , z) +
{\partial}_{z}^2 \left( \log < z \mid M_{+}^{-1} M_{-} \mid z >
\right) ~ ,
\ee
where ${\Psi}_{0} (q_{+} , q_{-} , z) = {\Psi}^{+} (q_{+} , z) +
{\Psi}^{-} (q_{-} , z)$ is the decomposition of the free field
configuration
into holomorphic and anti--holomorphic parts, and
\be
M_{\pm} (q_{\pm} , z) = {\rm P} \exp \left(\lambda \int^{q_{\pm}}
d q_{\pm}^{\prime} \int^{z} d z^{\prime}
e^{{\Psi}^{\pm} (q_{\pm}^{\prime} , z^{\prime})} X^{\pm}(z^{\prime})
\right) .
\ee
In this case all the terms (4.8) contribute for generic free field
configurations,
and the power series expansion of $< z \mid M_{+}^{-1} M_{-} \mid z
>$ is infinite
and could be divergent. Moreover, it is not clear at this point
whether every
solution of the continual Toda field equation can be obtained in this
fashion
or whether there exist solutions that do not admit a realization in
terms of
2--dim free fields according to (4.9).
We will see later that the Atiyah--Hitchin
metric precisely corresponds to such unusual solutions.

We analyse first the free field realization of the Toda potential
for the Eguchi--Hanson and Taub--NUT metrics.

(i) \underline{Eguchi--Hanson metric} : Introducing $q_{\pm} = (y \pm
ix)/2$
the asymptotic expansion of the Toda potential (3.10) is easily
obtained,
\be
e^{\Psi} = {2(z^2 - {\alpha}^2) \over q_{+}^2 q_{-}^2} \left(1 -
{4 \over q_{+} q_{-}} + {12 \over q_{+}^2 q_{-}^2} - {32 \over
q_{+}^3 q_{-}^3}
+ \cdots \right) .
\ee
Therefore, we may identify the free field configuration
${\Psi}^{+} (q_{+} , z) + {\Psi}^{-} (q_{-} , z)$ of the
Eguchi--Hanson metric choosing
\be
{\Psi}^{\pm} (q_{\pm} , z) = \log {\sqrt{2(z^{2} - {\alpha}^2)} \over
q_{\pm}^2} ~ .
\ee

It is fairly straightforward to verify that the expansion (4.11)
coincides with
(4.9) for this particular free field configuration, as it should. For
this
we compute first
\ba
& & D_{z}^{\{z_{1} ; z_{1}^{\prime}\}}  =
\delta (z, z_{1}) \delta (z_{1} , z_{1}^{\prime}) ~ , \nonumber \\
& & D_{z}^{\{z_{1} , z_{2} ; z_{1}^{\prime} , z_{2}^{\prime}\}}  =
\delta (z, z_{1}) \delta (z_{1} , z_{1}^{\prime}) \delta (z_{2} ,
z_{2}^{\prime})
\left(2 \delta (z_{1} , z_{2}) + {\partial}_{z_{1}}^2 {\delta} (z_{1}
, z_{2})
\right) ~ , \nonumber \\
& & D_{z}^{\{z_{1} , z_{2} , z_{3} ; z_{1}^{\prime} , z_{2}^{\prime}
,
z_{3}^{\prime}\}}
= \delta (z , z_{1}) \delta (z , z_{2}) \delta (z_{1} ,
z_{3}^{\prime})
\delta (z_{2} , z_{1}^{\prime}) \delta (z_{3} , z_{2}^{\prime})
\left(2 \delta (z_{1} , z_{3}) + {\partial}_{z_{1}}^2 \delta (z_{1} ,
z_{3})
\right) + \nonumber \\
& & ~~ \delta(z , z_{1}) \delta(z_{1} , z_{1}^{\prime})
\delta(z_{2} , z_{3}^{\prime}) \delta(z_{3} , z_{2}^{\prime})
\left( \delta (z_{1} , z_{2}) + {\partial}_{z_{1}}^2 \delta (z_{1} ,
z_{2}) \right)
\left(2 \delta (z_{1} , z_{3}) + {\partial}_{z_{1}}^2 \delta (z_{1} ,
z_{3})
\right) + \nonumber \\
& & ~~ \delta(z , z_{1}) \delta(z_{1} , z_{1}^{\prime})
\delta(z_{2} , z_{2}^{\prime}) \delta(z_{3} , z_{3}^{\prime})
\left(2 \delta (z_{1} , z_{2}) + {\partial}_{z_{1}}^2 \delta (z_{1} ,
z_{2}) \right)
\cdot \nonumber \\
& & ~~ \left(2 \delta (z_{1} , z_{3}) + {\partial}_{z_{1}}^2 \delta
(z_{1} , z_{3})
+ {\partial}_{z_{2}}^2 \delta (z_{2} , z_{3}) \right) ~ ,
\ea
etc. Then, using (4.12) we find the following result for the
expansion,
\ba
& & < z \mid M_{+}^{-1} M_{-} \mid z >
= 1 - (z^2 - {\alpha}^2) {2 \over q_{+} q_{-}} +
{1 \over 2} (z^{2} - {\alpha}^2) (z^2 - {\alpha}^2 + 1) \left({2
\over q_{+} q_{-}}
\right)^2 - \nonumber \\
& & {1 \over 6} (z^2 - {\alpha}^2) (z^2 - {\alpha}^2 + 1) (z^2 -
{\alpha}^2 + 2)
\left({2 \over q_{+} q_{-}}\right)^3 + \cdots ~ \equiv ~ {1 \over
\left(1 +
{2 \over q_{+} q_{-}} \right)^{z^2 - {\alpha}^2}} ~ ,
\ea
which indeed yields (4.11) when it is substituted into the
exponential form of
the master equation (4.9).

The Toda potential of the Eguchi--Hanson metric is special in that it
belongs
to the subclass of solutions of the continual Toda equation given by
an
ansatz with factorized $z$--dependence
\be
e^{\Psi (q_{+} , q_{-} , z)} = (z^2 + \beta z + \gamma)
e^{{\varphi}_{L} (q_{+} , q_{-})} ~ ,
\ee
provided that ${\varphi}_{L}$ solves the Liouville equation
\be
{\partial}_{+} {\partial}_{-} {\varphi}_{L} + 2 e^{{\varphi}_{L}} = 0
{}~ .
\ee
According to (4.7), and taking into account the sign difference that
appears
in the coupling constant between (4.6) (with $N=2$) and (4.16), we
write
down the general solution of the latter in free field realization,
\be
e^{{\varphi}_{L} (q_{+} , q_{-})} =
{e^{{\varphi}_{L}^{+} (q_{+}) + {\varphi}_{L}^{-} (q_{-})} \over
\left(1 + \int^{q_{+}} d q_{+}^{\prime}
e^{{\varphi}_{L}^{+} (q_{+}^{\prime})}
\int^{q_{-}} d q_{-}^{\prime} e^{{\varphi}_{L}^{-} (q_{-}^{\prime})}
\right)^2} ,
\ee
which is indeed described by two arbitrary functions of $q_{+}$ and
$q_{-}$,
namely
$\int^{q_{+}} d q_{+}^{\prime} e^{{\varphi}_{L}^{+}
(q_{+}^{\prime})}$
and
$\int^{q_{-}} d q_{-}^{\prime} e^{{\varphi}_{L}^{-}
(q_{-}^{\prime})}$
respectively. We find from this point of view that
\be
{\varphi}_{L}^{\pm} (q_{\pm}) = -2 \log q_{\pm} + \log \sqrt{2}
\ee
provides the relevant solution
that yields (4.12) using the ansatz (4.15) with
$\beta = 0$ and $\gamma = - {\alpha}^2$. This metric resides entirely
in the $SU(2)$ subalgebra of $SU(\infty)$, and hence is the simplest
one to
consider.

(ii) \underline{Taub--NUT metric} : The Toda potential of this metric
satisfies
the dimensionally reduced equation
\be
{\partial}_{y}^2 \Psi (y, z) + {\partial}_{z}^2 (e^{\Psi (y, z)}) = 0
{}~ ,
\ee
and therefore the asymptotic behaviour of the Toda field should be
linear in $y$,
\be
\Psi (y, z) \rightarrow \mp a(z) y + b(z) ~ ; ~~~~~ y \rightarrow \pm
\infty ~ ,
\ee
with $a(z) > 0$. We will determine the coefficients of the
corresponding free
field configuration by analysing the solution (3.19) in the limit
$y \rightarrow - \infty$, but a similar analysis can also be
performed around
$+ \infty$. As soon as $a(z)$ and $b(z)$ have been extracted, the
exact solution
will be described as a power series expansion in the usual way. The
analysis is
a little bit more involved than before, because
in this case the Toda potential is only
implicitly described as a function of $y$ and $z$. In the process
of analyzing the asymptotic form of $\Psi (y, z)$ a power
series expansion of $\theta$ and $t$ will also emerge
in terms of the $y$ and $z$ coordinates.
Throughout this process we keep $t$ arbitrary, but never equal to
zero although
it can be asymptotically close to it with exponentially small
corrections.

A close inspection of the equations (3.17) and (3.19) shows that the
limit
$y \rightarrow - \infty$ is achieved as $\theta \rightarrow 0$, in
which
case $\exp \Psi (y, z) \rightarrow 0$, as it should in the asymptotic
free field limit. In this limit we may approximate
$\theta \approx 2 e^{y+ z/m^2}$ and
$z \approx 1/4t$ to lowest order in $e^y$. Then, the Toda potential
approaches
zero with leading behaviour given by
$e^{\Psi} \approx 4z^2 e^{2y + 2z/m^2}$, and
consequently the asymptotic free field configuration (4.20) is given
by
\be
{\Psi}_{0} (y, z) = 2y + \log 4z^2 + {2z \over m^2} ~ .
\ee
Having determined $a(z) = 2$ and $b(z) = \log 4z^2 + 2z/m^2$
we may proceed in this case calculating
the form of the subleading terms of the expansion, in order to
compare them
with those following from the general solution of the continual Toda
equation.

It is convenient to introduce as an expansion parameter the following
quantity
\be
F(y,z) = e^{2y + 2z/m^2} \equiv {1 \over 4z^2} e^{{\Psi}_{0}} ~ ,
\ee
and use it for inverting the coordinate transformation
$(t, \theta) \rightarrow (y, z)$ in the asymptotic region. Explicit
calculation
yields the following result,
\be
\cos \theta = 1 - 2F + 2F^2 \left(1 + {2z \over m^2} \right)^2 - 2F^3
\left(1 + 12 {z \over m^2} + 48 {z^2 \over m^4} + 64 {z^3 \over m^6}
+ 24 {z^4 \over m^8} \right) + \cdots ~ ,
\ee
and
\be
{1 \over t} = 4z \left(1 - 2F {z \over m^2} + 4F^2 {z \over m^2}
\left(1+ {4z \over m^2} + {2z^2 \over m^4} \right) + \cdots \right) ,
\ee
where the dots denote higher order corrections in $F$. Using these
expansions
we also find that $\Psi (y, z)$ admits a power series expansion
around
the free field configuration (4.21),
\be
\Psi = {\Psi}_{0} - 2F \left(1 + {4z \over m^2} + {2z^2 \over m^4}
\right) +
F^2 \left(1 + 28 {z \over m^2} + 92 {z^2 \over m^4} + 96 {z^3 \over
m^6}
+ 30 {z^4 \over m^8} \right) + \cdots ~ .
\ee
It can be independently verified that this expression is consistent
with (4.19)
order by order in the expansion parameter $F$ (or ${\lambda}^2 F$, if
we had
supplied the non--linearity with a parameter ${\lambda}^2$ as in
(4.1)).

The free field (4.21) can be
applied in (4.9) and (4.10) in order to reformulate the
expansion of $\Psi(y, z)$ in a systematic way, as well as use it as
a cross--check
of the validity of
the series (4.25). For this purpose we have to find the general form
of the
solution of the continual Toda equation in the dimensionally reduced
sector,
where all the dependence is on $q_{+} + q_{-} = y$ and not on $q_{+}
- q_{-}$.
The correct way of doing this is to decompose ${\Psi}_{0}$ close to
$- \infty$
as
\be
{\Psi}_{0} (y, z) = {\Psi}^{+} (q_{+} , z) + {\Psi}^{-} (q_{-} , z) ~
; ~~~~~
{\Psi}^{\pm} = a(z) q_{\pm} + b_{\pm} (z) ~ ,
\ee
so that $b_{+} (z) + b_{-} (z) = b (z)$; in fact $b(z)$ is not
physically
important since it can be set equal to zero by shifting $y$. Then,
the
quantity $< z \mid M_{+}^{-1} M_{-} \mid z >$ in (4.9) can be
calculated
by expanding the path--ordered exponentials in a straightforward way
[9],
which for $a(z) = 2$ (independent of $z$) yields the result
\be
< z \mid M_{+}^{-1} M_{-} \mid z > = 1 - {1 \over 4} e^{{\Psi}_{0}} +
{1 \over 32} e^{2 {\Psi}_{0}} + {1 \over 64} e^{{\Psi}_{0}}
{\partial}_{z}^2 (e^{{\Psi}_{0}}) + \cdots ~ ,
\ee
where the dots denote third and higher order corrections in
$e^{{\Psi}_{0}} \equiv 4z^2 F$. Taking the logarithm and expanding it
in power series we may determine $\Psi (y, z)$ following the general
formula (4.9),
\be
\Psi = {\Psi}_{0} - {1 \over 4} {\partial}_{z}^2 (e^{{\Psi}_{0}})
+ {1 \over 64} {\partial}_{z}^2 \left(e^{{\Psi}_{0}} {\partial}_{z}^2
(e^{{\Psi}_{0}}) \right) + \cdots ~ .
\ee
This expansion is identical to (4.25), as it can be readily verified
using
the definition (4.22). It is systematically summarized by
$< z \mid M_{+}^{-1} M_{-} \mid z >$, which encodes all the
information for
$\Psi (y, z)$.

The Lie--Backlund transformation that relates the free field
configuration
${\Psi}_{0}$ with $\Psi$ can be regarded as a classical analogue
of the half--$S$--matrix connecting the asymptotic value of the field
at $- \infty$ with its value at $y$. From this point of view the
expansion in terms of free fields arises quite naturally
in the general theory of Toda systems.
One could have similarly considered
analyzing the asymptotic behaviour at $+ \infty$,
which for the Taub--NUT metric would lead to the free field
configuration
${\Psi}_{0} = -2y + \log 4z^2 +2z/m^2$.

Concluding this section we note that our results have a smooth limit
as the
moduli parameter of the Taub--NUT metric $m \rightarrow \infty$. The
coordinate transformations (4.23) and (4.24) contract to
$\cos \theta =  - \tanh y$ and $t = 1/4z$ respectively, in agreement
with (3.21).
Also the series (4.25) essentially provides the polynomial expansion
of
$1/ {\cosh}^2 y$ in powers of $e^{2y}$. We also note that the Toda
field
configuration in this case belongs to the special class of solutions
given by the ansatz (4.15), provided that the corresponding Liouville
field
is chosen to be
\be
e^{{\varphi}_{L} (y)} = {1 \over {\cosh}^2 y} ~ .
\ee
An alternative description of this is given by the formula (4.17)
making the
choice ${\varphi}_{L}^{\pm} (q_{\pm}) = 2 q_{\pm}$ as in (4.26).
Therefore,
in the limit $m \rightarrow \infty$ the solution can be thought as
residing
entirely in the $SU(2)$
subalgebra of $SU(\infty)$, and it ``spreads" into the whole of
$SU(\infty)$
by including all the $1/m^2$ corrections. Furthermore, close to the
nut
located at $r=m$ (or equivalently at $t \rightarrow \infty$)
we find that the Toda potential of the Taub--NUT metric
(with finite $m$) is also given by
$e^{\Psi} \approx z^2 / {\cosh}^2 y$; in this case, however, $z$
approaches zero
and the complete description of the metric requires
the power series corrections
in $z$ appearing in the expansion (4.25). These two remarks complete
our
understanding of the way that the full $SU(\infty)$ algebra is
employed for
the description of this metric.

\section{\bf Analysis of the Atiyah-Hitchin metric}
\setcounter{equation}{0}
\noindent
Searching for the free field realization of a given metric requires
first
to locate the zeros of $e^{\Psi}$. Straightforward calculation
shows that for the Atiyah--Hitchin metric, and for generic values
of $0 < k < 1$, the zeros of (3.27) occur at
\be
{\cos}^2 \theta = k^2 ~ ; ~~~~ \psi = {\pi \over 2} ~~ {\rm or} ~~
{3 \pi \over 2} ~.
\ee
These four points are actually not distinct from each other,
due to certain identifications imposed by discrete symmetries
on the angle variables $0 \leq \theta \leq \pi$, $0 \leq \psi \leq 2
\pi$
and $0 \leq \phi \leq 2 \pi$. Explicitly, we consider the symmetry
operations
\be
\theta \rightarrow \pi - \theta ~ , ~~~~ \psi \rightarrow 2 \pi
- \psi ~, ~~~~ \phi \rightarrow \pi + \phi ~ ,
\ee
and
\be
\theta \rightarrow \theta ~ , ~~~~ \psi \rightarrow \pi + \psi ~,
{}~~~~
\phi \rightarrow \phi ~,
\ee
which are clearly invariances of ${\sigma}_{1}^2$, ${\sigma}_{2}^2$
and ${\sigma}_{3}^2$. Note that under (5.2), which exchanges the
position of
monopoles so that they are treated as identical particles, the point
$(\cos \theta , \psi) = (k , ~ \pi / 2)$ is mapped to
$(-k , ~ 3 \pi / 2)$, and $(k , ~ 3 \pi / 2)$ is mapped to
$(-k , ~ \pi / 2)$. Similarly, under (5.3) the point $(k , ~ \pi /
2)$ is mapped
to $(k , ~ 3 \pi / 2)$, and $(-k , ~ \pi / 2)$ is
mapped to $(-k , ~ 3 \pi / 2)$. The discrete transformations
(5.2) and (5.3) are invariances of the Toda potential (3.27),
but they have the following effect on the coordinates $\nu$,
given by (3.24), and $q = y + i x$, given by (3.25):
\be
\nu \rightarrow - \nu + 4 \pi i ~ , ~~~~ q \rightarrow -q
\ee
\be
\nu \rightarrow \nu + 2 \pi i ~ , ~~~~ q \rightarrow q
\ee
respectively. In the Atiyah--Hitchin metric we always assume
the identifications imposed by (5.2), and hence there is a folding
(5.4) that results to a cone--like structure around $q = 0$, but
eventually we also impose (5.3) when considering the metric on
the moduli space of $SU(2)$ BPS monopoles $M_{2}^{0}$ (for details
see [5]).
Thus, up to a double covering implied by (5.3), it suffices
to consider
\be
\cos \theta = k ~ , ~~~~ \psi = {\pi \over 2}
\ee
as describing the unique free field point,
and let
$0 \leq \theta \leq \pi / 2$ and $0 \leq \psi \leq \pi$.

The moduli space $M_{2}^{0} = {\tilde M}_{2}^{0} / Z_{2}$ has
non--trivial
homotopy ${\pi}_{1} (M_{2}^{0}) = Z_{2}$, and hence its double
covering
${\tilde M}_{2}^{0}$ is simply connected and universal [5]. In the
$\nu$--plane there are two free field points, say $Q_{0}^{\pm}$ that
correspond to $(\cos \theta , ~ \psi) = (k , ~ \pi / 2)$ and
$(k , ~ 3 \pi / 2)$ in the center of the ``in" and ``out" regions,
while
$q$ and $e^{\Psi}$ are invariant under the $Z_{2}$ symmetry (5.3)
and so they are well--defined on the monopole moduli space.
Our analysis essentially takes place in the simply connected space
${\tilde M}_{2}^{0}$, choosing to work with (5.6). We assume
in the following the free
field realization of Toda field configurations as in flat space.
The multi--valuedness of the transformation (3.25) for $\theta = 0$,
which yields $q = 0$ for all values of the angle variable $\psi$
(also
$\Psi$ is independent of $\psi$ for $\theta = 0$),
will be regarded as a coordinate
singularity analogous to the origin of a spherical
coordinate system. We also note that at the free field point (5.6)
the
coordinates $\theta$ and $k$ are related to each other, whereas in
the
Taub--NUT limit we chose earlier $\theta = 0$ independently of the
other coordinates. Hence, we do not expect the free field analysis of
the Atiyah--Hitchin metric to yield smoothly the Taub--NUT free
field configuration for $k \rightarrow 1$, although the line
elements themselves become exponentially close to each other.

In principle, we could have considered points where
$e^{\Psi} \rightarrow az + b$, assuming that $a$ and $b$ are
$z$--independent
constants, since the non--linear term ${\partial}_{z}^2 e^{\Psi}$
would have also been zero there. A typical example that illustrates
the
relevance of this possibility is the solution $e^{\Psi} = z/2$,
independent
of $q$ and $\bar{q}$ everywhere, which describes the flat space
metric
in polar coordinates after a suitable change of variables. The Toda
field is identical to the free field in this case, and a simple
calculation shows that the logarithmic term in (4.9) is linear in
$z$,
thus giving zero contribution when ${\partial}_{z}^2$ acts on it.
We have not considered this possibility for the Atiyah--Hitchin
metric
because there already exists the point (5.6) for which $e^{\Psi} =
0$.

Next, in analogy with the analysis done for the Taub--NUT metric,
we have to invert the transformation (3.25) in the vicinity of
the free field point (5.6), and try to express directly
the corresponding
Toda potential in terms of $q$ and $\bar{q}$. It is convenient
to parametrize this region by two real variables ${\epsilon}_1$
and ${\epsilon}_2$, so that
\be
\cos \theta = k - {1 \over 2} {k^{\prime}}^2 {\epsilon}_{1} ~ ,
{}~~~ \psi = {\pi \over 2} + {1 \over 2} {\epsilon}_{2} ~ ,
\ee
and also introduce the complex variable
$\epsilon = {\epsilon}_{1} + i {\epsilon}_{2}$. Then, to lowest
order in $\epsilon$, we have
\be
\nu = {\nu}_{0} + \epsilon ~ ; ~~~~
{\nu}_{0} = \log {1 - k \over 1 + k} + i \pi ~ ,
\ee
and
\be
1 + {k^{\prime}}^2 {\sinh}^2 {\nu \over 2} = k \epsilon ~ , ~~~
k \tanh {\nu \over 2} = - \left(1 + {{k^{\prime}}^2 \over 2k}
\epsilon \right) ~ .
\ee
Therefore, the Toda potential admits around ${\nu}_{0}$ the
expansion
\be
e^{\Psi} = {1 \over 16} k {k^{\prime}}^2 {K}^2 (k)
\mid \epsilon \mid + {\cal O} ({\epsilon}^2) ~ .
\ee
We also find that the coordinate $z$ given by (3.23) has the
expansion
\be
z = {1 \over 4} K (k) ({k^{\prime}}^2 K (k) - E (k)) + {\cal O}
(\epsilon) ~ .
\ee

The expansion of the coordinate $q$ around ${\nu}_{0}$ is a bit
trickier because the elliptic integral in (3.25) diverges as
$\epsilon \rightarrow 0$, while
$\sqrt{1 + {k^{\prime}}^2 {\sinh}^2 \nu / 2}$ tends to zero.
It is convenient to rewrite
\be
\int_{0}^{\pi / 2} d \gamma {\sqrt{1 - k^2 {\sin}^2 \gamma} \over
1 - k^2 {\tanh}^2 {\nu \over 2} {\sin}^2 \gamma} =
{K(k) \over {\tanh}^2 {\nu \over 2}} - {1 \over {\sinh}^2
{\nu \over 2}} \Pi (k^2 {\tanh}^2 {\nu \over 2} , ~ k) ~ ,
\ee
where
\be
\Pi (n^2 , k) = \int_{0}^{\pi / 2} {d \gamma \over (1 - n^2
{\sin}^2 \gamma) \sqrt{1 - k^2 {\sin}^2 \gamma}}
\ee
is the complete elliptic integral of the third kind, and study
its expansion close to $\mid n \mid = 1$. Note that (5.13) is
well--defined for $\mid n \mid < 1$, while it diverges on the
real axis for all $\mid n \mid \geq 1$, and the Cauchy principal
value of the integral representation has to be taken into account
there [16]. In fact,
for $\psi = \pi / 2$, $n^2 < 1$
corresponds to $\cos \theta < k$, and $n^2 \geq 1$ to
$\cos \theta \geq k$.

The correct definition of the complete elliptic integral of the third
kind, for all complex values of $n$, is provided by the analytic
continuation in terms of theta functions.
We parametrize $n^2 = k^2 {\rm sn}^2 \alpha$, using the Jacobi
elliptic
function of a complex variable $\alpha$, and arrive at the
expression [16]
\be
\Pi (k^2 {\rm sn}^2 \alpha, ~ k) = K(k) + {\pi \over 2 k^{\prime}}
{{\theta}_{4}^{\prime} \left({\pi \alpha \over 2 K(k)} \right)
{\theta}_{1} \left({\pi \alpha \over 2 K(k)} \right) \over
{\theta}_{2} \left({\pi \alpha \over 2K(k)} \right)
{\theta}_{3} \left({\pi \alpha \over 2K(k)} \right)} ~ ,
\ee
where the modulus of the theta functions is $i K(k^{\prime}) / K(k)$,
which is proportional to the coordinate variable $t$ in (2.19).
Then, (5.13) admits the following expansion
\be
\Pi (n^2 , k) = K(k) - {E(k) \over {k^{\prime}}^2} +
{\pi (1 + {k^{\prime}}^2 - k^2 n^2) \over 4 {k^{\prime}}^3
\sqrt{1 - n^2}} + {\cal O} (1 - n^2) ~ ,
\ee
which yields
\be
\Pi (k^2 {\tanh}^2 {\nu \over 2} , ~ k) = K(k) - {E(k) \over
{k^{\prime}}^2} + {\pi \sqrt{k} \over 2 {k^{\prime}}^2
\sqrt{- \epsilon}} + {\cal O} (\sqrt{- \epsilon})
\ee
using the parametrization (5.9) around ${\nu}_{0}$.
This result, which is actually valid
for all complex values of the parameters, leads to the expansion
of the coordinate $q$
\be
q = q_{0} + {E(k) - {k^{\prime}}^2 K(k) \over \sqrt{k}}
\sqrt{\epsilon}
+ {\cal O} (\epsilon) ~ ,
\ee
where $q_{0} = i \pi / 2$. There is an overall sign ambiguity in
evaluating $q$ that depends on which branch we choose for
$\sqrt{-1} = \pm i$. However, thanks to the discrete symmetry (5.4)
that has been imposed on the coordinate system of the Atiyah--Hitchin
metric this ambiguity is not relevant. Putting everything together
we obtain the desired expansion of the Toda potential
\be
e^{\Psi} = {1 \over 16} \left({k k^{\prime} K(k) \over
E(k) - {k^{\prime}}^2 K(k)} \right)^2 {\mid q - q_{0} \mid}^2 +
{\cal O} ({\mid q - q_{0} \mid}^4) ~ .
\ee
The $k$-dependent coefficient of ${\mid q - q_{0} \mid}^2$ term
can be expressed as a function of $z$,
say $e^{f(z)}$, using (5.11) to lowest order,
although there is no closed formula for it.

The main point of this analysis is to demonstrate explicitly that
there is no free field realization of the Atiyah--Hitchin metric
along the lines of the previous section. Indeed, if we consider
\be
{\Psi}_{0} = \log {\mid q - q_{0} \mid}^2 + f (z)
\ee
as the corresponding free field extracted from the expansion (5.18),
then the path--ordered exponentials (4.10) will
depend only on the combination $\mid q - q_{0} \mid$, and so
the complete Toda potential (4.9) will depend only on
$\mid q - q_{0} \mid$ as well.
We realize, therefore,
that if the Toda potential
of the Atiyah--Hitchin metric were to admit a free field realization
according to (4.9), the metric would exhibit an additional $U(1)$
isometry, which is manifest in $\mid q - q_{0} \mid$, and hence
commute with $SO(3)$. Since this is not geometrically allowed,
we conclude that this metric corresponds to a new class of
solutions of the continual Toda equation that can accomodate
the breaking of the additional
$U(1)$ isometry. The essential ingredient here is the factorization
of the coordinate variables $q$ and $z$ implied by (5.19) in the
``would be" free field configuration. If such a factorization were
not
present, it would have been impossible to translate the reality
condition of the corresponding ${\Psi}_{0}$ into a manifest $U(1)$
isometry of the metric by suitable conformal transformation of the
coordinates
$q$ and $\bar{q}$; in Liouville theory the reality condition of the
solutions in Euclidean space always yields a manifest
$U(1)$ isometry, but in the continual Toda theory the situation is
different, unless the $z$--dependence factorizes,
because of the operator ${\partial}_{z}^2$ that acts on $e^{\Psi}$.

We have no systematic description of such new
solutions generalizing (4.9). One may think that there is a possible
resolution by introducing suitable prefactors in (4.10). The
particular
choice that was made for (4.10) corresponds to some fixed boundary
condition, but in general we could have integrated the continual
analogue of the linearized equations (4.4) to
\be
\tilde{M}_{\pm} = M_{\pm}^{(0)} {\rm P} \exp \left( \lambda
\int^{q_{\pm}} dq_{\pm}^{\prime} \int^{z} dz^{\prime}
e^{{\Psi}^{\pm} (q_{\pm}^{\prime} , z^{\prime})} X^{\pm} (z^{\prime})
\right)
\ee
with constant group elements $M_{\pm}^{(0)}$. In doing that we should
consider
solutions (4.9) with
\be
< z \mid \tilde{M}_+^{-1} \tilde{M}_- \mid z > =
< z \mid M_+^{-1} g_0^{(0)} M_- \mid z > ~,
\ee
where
\be
g_{0}^{(0)} = (M_+^{(0)})^{-1} M_-^{(0)} ~.
\ee
So the solutions are characterized in reality by the free field
specialization
as well as the choice of a constant group element $g_0^{(0)}$. This
point is
very important in affine Toda theory because soliton solutions have
zero free fields, but non--trivial $g_0^{(0)}$ [21]. We have analysed
whether
this modification can change the previous conclusions. We have
verified
that there is no consistent choice of $g_0^{(0)}$, which is the
exponential
of Lie algebra elements,
that breaks the unwanted $U(1)$ isometry while preserving the reality
of
the metric.

We think that responsible for this situation
are held the topological properties of the group of area preserving
diffeomorphisms,
whose algebra describe $SU(N)$ in the continual large $N$ limit. In
ordinary Toda
theory one uses the fact that the exponential of an algebra is a Lie
group that
admits the Gauss decomposition [12]. For $SU(\infty)$, however, the
exponentiation of the algebra of diffeomorphisms is not a Lie group
in the
topological sense (see for instance [22]), and hence the existence of
a
Gauss type decomposition is a topological assumption that one makes
to
arrive eventually to (4.9).

Since the Atiyah--Hitchin metric provides
the first example of such
non--trivial configurations in Toda theory, it deserves further
study in view of possible generalizations in rotational
hyper--Kahler geometry. It is rather remarkable that the purely
rotational character of this metric, due to the
absence of an additional $U(1)$ translational isometry, has
profound implications in monopole scattering, where slowly moving
monopoles can be converted into dyons [5]. The absence of
this isometry is also responsible for the special
features of the corresponding Toda field configuration, but we
do not understand very well whether there exists a deeper relation
between the
particle--like approximation of monopole scattering (given by the
Taub--NUT limit of the metric) and the possibility to have
free field realizations of the Toda potential. Also, the discrete
$Z_{2}$
symmetry (5.3) has no effect on the presence or the absense of an
additional continuous $U(1)$ isometry in the metric, and so our
analysis should
be insensitive to it.

\section{\bf Conclusions and discussion}
\setcounter{equation}{0}
\noindent
We have presented a systematic study of the Toda field formulation
of 4--dim hyper--Kahler metrics with a rotational isometry.
The only examples known to this date are the Eguchi--Hanson,
Taub--NUT and Atiyah--Hitchin metrics, which actually exhibit a
bigger group of isometries containing $SO(3)$.
We found that the qualitative differences between them
can be better understood in algebraic terms by considering the
free field realization of the corresponding Toda field
configurations. The solutions of the continual
Toda equation that were obtained by taking the large $N$
limit of the $SU(N)$ Toda theory seem to be inadequate
for describing the Atiyah--Hitchin metric. For the
other two metrics, however, the 2-dim free field specialization of
Saveliev's group theoretical formula exists and it has been
given explicitly.

The Atiyah--Hitchin metric can be regarded as
the simplest representative from a class of 4--dim purely
rotational hyper--Kahler metrics. We note, however, that such a
generalized series of regular metrics (if they exist) can
have only one rotational isometry, and
no additional isometries of either type.
Recall the observation of Boyer and Finley that there are no real
hyper--Kahler manifolds with two rotational isometries whose
algebra closes upon itself [7],
\be
[K_{1} , ~ K_{2}] = \alpha K_{1} + \beta K_{2} ~ .
\ee
If one insists to have two isometries satisfying (6.1) the structure
constants can be chosen without loss of generality so that $\alpha =
0$
and $\beta$ is purely imaginary, and the
self--duality condition of the metric forces one of the
Killing isometries to be translational. Moreover, it can also be
easily shown that the generators of rotational isometries
either form an $SO(3)$ algebra or there is only one of them, since
otherwise their algebra would contain a two--dimensional solvable
subalgebra. According to the classification of $SO(3)$
hyper--Kahler metrics only the Atiyah--Hitchin metric defines
a complete and regular space of the first kind, with no additional
translational isometries, and as a result we expect that
a descending series of purely rotational metrics in
four dimensions will necessarily exhibit only one isometry. Hence, it
might be difficult to obtain candidate new metrics in
closed form.

We mention finally that certain generalizations of the
Atiyah--Hitchin metric have been already proposed [5],
using a different method.
The moduli space of BPS $SU(2)$ monopoles of magnetic charge 2 is
associated to the space of rational functions
\be
f(z) = {u z + v \over z^2 + w} ~ ; ~~~~ v^2 + w u^2 \neq 0 ~ ,
\ee
and the equation $v^2 + w u^2 = 1$ defines a manifold in $C^3$ which
is a double covering of it. This equation can be regarded as a
special
case of the more general equation $v^2 + w u^2 = w^{k-1}$ in $C^3$,
which leads to a series of hyper--Kahler metrics associated to the
orthogonal groups $SO(2k)$ (D--series). This algebraic description
is analogous to the A--series generalization of the Taub--NUT
metric, where the multi--Taub--NUT solutions have a characteristic
equation $uv = w^k$ related to $SU(k)$ (for details on polygon
constructions see [6]). Recently, there has been a discussion
of this in the literature, using the twistor space interpretation
of the Legendre transform construction of hyper--Kahler metrics
[23]. In that regard there are certain similarities with the
Toda frame formulation of rotationally invariant metrics.
Our approach, however, is quite different
emphasizing the presence of a rotational isometry in the (proposed)
series of metrics that descend from Atiyah--Hitchin; whereas in the
D--series generalization, it is unlike that any isometries will be
at all present.

In any case, it will be interesting to explore further all these
possibilities, and answer the question whether the
Atiyah--Hitchin metric is the unique purely rotational complete
hyper--Kahler 4--metric or it is indeed the first representative from
a new series of metrics. One drawback at this stage is the
local nature of our method, which makes rather delicate the issue
of the completeness of the metrics that can be obtained from
different
solutions of the continual Toda equation.

\vskip .6in
\centerline{\bf Acknowledgements}
\vskip.2cm
\noindent
One of us (K.S.) wishes to thank the Theory Division of CERN
for hospitality during the course of this work. We also thank
Elias Kiritsis and Misha Saveliev for useful remarks on the
manuscript.
The work of K.S. was carried out with the financial support of the
European Union research program ``Training and Mobility of
Researchers"
and is part of the project ``String Duality Symmetries and Theories
with Space--Time Supersymmetry".

\newpage

\centerline{\bf Note Added}

An interesting series of papers came recently to our attention,
which prove the existence of at least two families
of new non--singular 4--dim hyper--Kahler manifolds with 
rotational $SO(2)$ isometry. 

The first family describes a 1--parameter deformation of the 
Atiyah--Hitchin space as complex hypersurface in $C^3$,
$v^2 + wu^2 = 1 + i m u$.
The 4-dim metric was implicitly described by Dancer [24]
starting from a 12--dim moduli space of solutions to Nahm's
equations and then performing suitable hyper--Kahler 
quotients of it. For $m = 0$ we obtain the Atiyah--Hitchin 
space, but for $m \neq 0$ the resulting manifold has only 
one purely rotational isometry. It is interesting that these
spaces arise naturally as moduli spaces of the 
supersymmetric 3--dim $SU(2)$ gauge theory with $N_{f} = 1$
and a bare mass $\sim m$ [25]. 

The second family corresponds to the hypersurface in $C^3$,
$v^2 + wu^2 = w + ikuv + lv$,
which is non-singular if and only if $k \neq \pm l$ [26]. 
It also turns out that the resulting 4--dim hyper--Kahler
manifolds exhibit no isometries, with the only exception of 
${\rm Im}l = 0$ that gives rise to spaces with one
rotational isometry. A more intuitive geometric construction
was apparently proposed by Page a number of years ago through
a limiting procedure from the Einstein metric on $K_3$ [27].
In this case we have a 3--parameter family of purely 
rotational hyper--Kahler manifolds although the explicit
form of the metric is also not known. It is conceivable 
that suitable deformations of the more general hypersurfaces
$v^2 + wu^2 = w^{k-1}$ may exist as well that yield purely
rotational hyper--Kahler manifolds for appropriate limits of
the relevant deformation parameters. 

These additional results supplement nicely our point of view,
thus making more urgent the question of the explicit metrics
and the construction of the corresponding Toda field 
configurations. Perhaps there is a more direct relation
between $SU(\infty)$ Toda theory, moduli spaces of Nahm's
equations and the deformations of hypersurfaces in $C^3$
that were considered above.

\newpage
\centerline{\bf REFERENCES}
\begin{enumerate}
\item G. Gibbons and S. Hawking, Commun. Math. Phys. \underline{66}
(1979) 291;
T. Eguchi, P. Gilkey and A. Hanson, Phys. Rep. \underline{66} (1980)
213, and
references therein.
\item R. Penrose, Gen. Rel. Grav. \underline{7} (1976) 31.
\item M. Ko, M. Ludvigsen, E.T. Newman and K. Tod, Phys. Rep.
\underline{71}
(1981) 51, and references therein.
\item L. Alvarez--Gaume and D. Freedman, Commun. Math. Phys.
\underline{80}
(1981) 443; N. Hitchin, A. Karlhede, U. Lindstrom and M. Rocek,
Commun. Math. Phys.
\underline{108} (1987) 535.
\item M. Atiyah and N. Hitchin, Phys. Lett. \underline{A107} (1985)
21;
``The Geometry and Dynamics of Magnetic Monopoles", Princeton
University
Press, 1988; G. Gibbons and N. Manton, Nucl. Phys. \underline{B274}
(1986) 183.
\item N. Hitchin, Math. Proc. Camb. Phil. Soc. \underline{85} (1979)
465;
P. Kronheimer, J. Diff. Geom. \underline{29} (1989) 665, 685.
\item C. Boyer and J. Finley, J. Math. Phys. \underline{23} (1982)
1126;
J. Gegenberg and A. Das, Gen. Rel. Grav. \underline{16} (1984) 817.
\item G. Gibbons and P. Rubback, Commun. Math. Phys. \underline{115}
(1988) 267.
\item M. Saveliev, Commun. Math. Phys. \underline{121} (1989) 283;
R. Kashaev, M. Saveliev, S. Savelieva and A. Vershik, in ``Ideas and
Methods in
Mathematical Analysis", volume in memory of R. Hoegh--Krohn,
Cambridge
University Press, 1992; M. Saveliev, Theor. Math. Phys.
\underline{92} (1993) 1024.
\item I. Bakas, in ``Supermembranes and Physics in $2+1$ Dimensions",
eds.
M. Duff, C. Pope and E. Sezgin, World Scientific, Singapore, 1990;
Commun. Math. Phys. \underline{134} (1990) 487; Q.--H. Park, Phys.
Lett.
\underline{B238} (1990) 287.
\item G. Gibbons and C. Pope, Commun. Math. Phys. \underline{66}
(1979) 267.
\item A. Leznov and M. Saveliev, Lett. Math. Phys. \underline{3}
(1979) 489;
Commun. Math. Phys. \underline{74} (1980) 111; P. Mansfield, Nucl.
Phys.
\underline{B208} (1982) 277.
\item J. Hoppe, Phys. Lett. \underline{B215} (1988) 706; D. Fairlie
and
C. Zachos, Phys. Lett. \underline{B224} (1989) 101; E. Floratos,
Phys. Lett. \underline{B232} (1989) 467.
\item K. Sfetsos, Phys. Rev. \underline{D54} (1996) 1682.
\item D. Olivier, Gen. Rel. Grav. \underline{23} (1991) 1349.
\item P. Byrd and M. Friedman, ``Handbook of Elliptic Integrals for
Engineers and Physicists", Springer--Verlag, 1954; E. Whittaker and
G. Watson, ``A Course of Modern Analysis", Cambridge University
Press, 1969.
\item V. Belinskii, G. Gibbons, D. Page and C. Pope, Phys. Lett.
\underline{B76} (1978) 433; G. Gibbons, D. Olivier, P. Rubback and
G. Valent, Nucl. Phys. \underline{B296} (1988) 679.
\item I. Bakas and K. Sfetsos, Phys. Lett. \underline{B349} (1995)
448.
\item M. Prasad, Phys. Lett. \underline{B83} (1979) 310.
\item M. Sakaguchi, Int. J. Mod. Phys. \underline{A11} (1996) 1279.
\item D. Olive, N. Turok and J. Underwood, Nucl. Phys.
\underline{B401} (1993) 663; ibid \underline{B409} (1993) [FS] 509.
\item M. Adams, T. Ratiu and R. Schmid, in ``Infinite Dimensional
Groups with
Applications", ed. V. Kac, Spinger--Verlag, 1985.
\item I. Ivanov and M. Rocek, ``Supersymmetric $\sigma$--models,
Twistors, and the Atiyah--Hitchin Metric", Stony Brook preprint
ITP--SB--95--54, hep--th/9512075, December 1995.
\item A. Dancer, Comm. Math. Phys. \underline{158} (1993)
545; ``A Family of Hyper--Kahler Manifolds", Cambridge
preprint DAMTP 91--41.
\item N. Seiberg and E. Witten, ``Gauge Dynamics and 
Compactification to Three Dimensions", hep--th/9607163.
\item A. Dancer, ``A Family of Gravitational Instantons",
Cambridge preprint DAMTP 92--13.
\item D. Page, Phys. Lett. \underline{B100} (1981) 313.
\end{enumerate}
\end{document}